# On the origin of perpendicular magnetic anisotropy in strained Fe-Co(-X) films

*L. Reichel[1,2], A. Edström[3], D. Pohl[1], J. Rusz[3], O. Eriksson[3], L. Schultz[1,2], and S. Fähler[1*]*

[1] IFW Dresden, P.O. Box 270116, 01171 Dresden, Germany

[2] TU Dresden, Faculty of Mechanical Engineering, Institute of Materials Science, 01062 Dresden, Germany

[3] Division of Materials Theory, Department of Physics and Astronomy, Uppsala University, Box 516, SE-751 20, Uppsala, Sweden

*corresponding author

**Abstract**

Very high magnetic anisotropies have been theoretically predicted for strained Fe-Co(-X) and indeed several experiments on epitaxial thin films seemed to confirm strain induced anisotropy enhancement. This study presents a critical analysis of the different contributions to perpendicular anisotropy: volume, interface and surface anisotropies. Tracing these contributions, thickness series of single layer films as well as multilayers with Au-Cu buffers/interlayers of different lattice parameters have been prepared. The analysis of their magnetic anisotropy reveals a negligible influence of the lattice parameter of the buffer. Electronic effects, originating from both, the Au-Cu interface and the film surface, outrange the elastic effects. Surface anisotropy, however, exceeds the interface anisotropy by more than a factor of three. A comparison with results from Density Functional Theory suggests, that the experimentally observed strong perpendicular surface anisotropy originates from a deviation from an ideal oxide-free surface. Accordingly, tailored Fe-Co-X/oxide interfaces may open a route towards high anisotropy in rare-earth free materials.

PACS:

75.30.Gw     Magnetic anisotropy

75.70.Ak     Magnetic properties of monolayers and thin films

75.50.Bb     Fe and its alloys

75.50.Ww    Permanent magnets

**Motivation**

Nearly one hundred years ago, polycrystalline Co steels were state of the art permanent magnets[1,2]. With the rise of Al-Ni-Co and the rare earth based alloys, the interest on these so called KS or Honda steels was suppressed. Within the last decade, new interest has risen because new methods of computation allow for a deeper understanding of magnetic anisotropy in single crystalline Fe-Co. Thin films of Fe-Co are now in the focus of research since large magnetocrystalline anisotropies were predicted, when their lattice is strained tetragonally[3-5]. A common approach to achieve such strain is coherent growth on substrates or buffer layers with certain lattice parameters, which apply compressive stress in the films' plane. While the in-plane lattice parameter *a* is compressed, the stress-free perpendicular axis *c* expands in order to preserve the volume of the Fe-Co unit cells. Following the concept of the epitaxial Bain path (EPB)[6], many different buffers and their influence on the strain and the magnetic properties of Fe-Co were investigated including Pt[7,8], Pd[9], Ir[10,11], Rh[12,13] and $Au_xCu_{1-x}$[14]. The latter alloyed buffer allows for a continuous variation of lattice parameters[15] as they depend on the stoichiometry x. In films with thicknesses of few nanometers, strong perpendicular magnetic anisotropies were observed, which has been interpreted as a confirmation of the named theoretical calculations[3-5]. For thicker films, however, a complete relaxation of the buffer induced strain is expected and observed[16] due to the significant elastic energy of strained Fe-Co, which was calculated by Neise et al.[4]. With increasing thickness of the Fe-Co films degrading magnetic anisotropy is thus reported[13,14,17]. Magnetic anisotropy is commonly measured for the complete film's volume, i.e. with integral methods, and thus it is challenging to distinguish between thickness dependent and independent contributions, which is required to separate between strain related anisotropy and volume anisotropy. While the latter is not dependent on film thickness, the induced strain originates from the buffer interface and its contribution to the magnetic anisotropy should be considered as thickness dependent[18]. This dependency was not accounted sufficiently in several studies[7,12], which intended to determine interface or surface anisotropies of thin Fe-Co films.

As lattice relaxation already proceeds at film thicknesses of few monolayers, the determination of the present strain is also very difficult and sensitive to systematic errors. Electron diffraction studies were often applied to study the lattice strain[9,10,12-14] of the deposited films. Strain ratios (*c/a*) were often indirectly calculated from measurements of the in-plane lattice parameters *a* with Low Energy Electron Diffraction (LEED)[9,10,12] and Reflection High Energy Electron Diffraction (RHEED)[13], just assuming the Fe-Co unit cells to preserve their volume without measuring out-of-plane lattice parameters *c* directly. X-ray diffraction to characterize the present strain accurately is also hardly suitable since the intensity of diffracted radiation is very low for ultrathin films[16,19]. These experimental limits might explain, why e.g. *c/a* ratios up to 1.32 were reported for Fe-Co films on Pt[8,20], although a strain ratio around 1.08 would be expected for volume preserving deposition.

With the aim to stabilize tetragonal strain (*c/a* > 1) in Fe-Co at higher film thicknesses, additions of small atoms like B or C were proposed[21,22]. Such atoms occupying octahedral interstitial sites in the Fe-Co lattice indeed result in a tetragonal strain of the Fe-Co lattice, which support a magnetocrystalline anisotropy (MCA) with its easy axis parallel to the strained *c* axis[14,21]. As such spontaneous lattice strain is independent from the films' thicknesses and is present up to thicknesses of at least 100 nm, the corresponding magnetic anisotropy is considered to be constant within the films' volume. Their uniaxial anisotropy constant is in the order of 0.4 $MJ/m^3$ [14,21]. However, the perpendicular anisotropy of much thinner $(Fe_{0.4}Co_{0.6})_{0.98}C_{0.02}$ films with a thickness of 5 nm is in the range of 0.8 $MJ/m^3$, which is twice the value of this volume anisotropy, though their tetragonal strain is in the same range[17]. This comparison already indicates that a high thickness dependent contribution to the anisotropy should be considered, even for film thicknesses, at which all buffer induced strain is relaxed.

From this short review of experimental results, important open questions arise, which are decisive for the understanding of thin Fe-Co films: Which property of an interface determines the magnetic anisotropy: induced lattice strain or other, electronic, contributions? How large are these contributions and which one dominates? Is it possible to distinguish contributions from the buffer interface and the film's surface? And, finally: Do these results confirm the initial idea of strain induced anisotropies? In our study, we compare single layered Fe-Co-X films with multilayers on buffers and interlayers with different lattice parameters. These experiments, together with DFT calculations, allow for a separation of elastic and electronic contributions to anisotropy and quantify the role of the buffer interface and the upper free surface. Our analysis shows that the major contribution to the perpendicular magnetic anisotropy originates from the free surface, which is presumably oxidized.

**Experimental and theoretical methods**

Film preparations were performed exploiting Pulsed Laser Deposition (PLD) in ultra-high vacuum ($6 \times 10^{-9}$ mbar) at room temperature. All Fe-Co-X films were deposited on MgO(100) with a 3 nm thick Cr buffer layer and 30 nm thick Au-Cu buffer layer. The composition of the latter has been varied within the present study as this allows for a variation of the buffer lattice parameters[15,17]. For the presented multilayer films, 4 nm thick Au-Cu with the same composition like the buffer was deposited on top of the Fe-Co-X films. This sequence, Fe-Co-X/Au-Cu, was repeated $n$ times. For these multilayer films, the top layer was Au-Cu. In contrast, no additional protection layer was used for the Fe-Co-X single layer films. Accordingly, one can expect the formation of a thin native oxide before *ex situ* measurements were performed. Further details about the depositions, such as the used parameters and targets, were introduced in our previous works[14,21]. The film compositions were confirmed with Energy Dispersive X-ray spectroscopy (EDX) measurements in a JEOL JSM6510-NX scanning electron microscope. The concentrations of B or C in the studied Fe-Co-X films have been determined with Auger Electron Spectroscopy (AES) on similar films in our previous studies[14,21], where identical deposition parameters were applied. We thus assume their concentration to be constant in all presented films.

XRD studies have been conducted in Bragg-Brentano geometry on a Bruker D8 Advance with CoKα source and on an X'Pert four circle goniometer using CuKα radiation. X-ray reflectivity (XRR) measurements were performed on a Panalytical X'Pert Pro PW3040/60, also using CuKα, and simulated with the software X'Pert Reflectivity 1.3a.

Transmission Electron Microscopy (TEM) was carried out on a Titan³ 80-300 microscope with a Schottky field emission electron source and a $C_S$ corrector.

Magnetic ex situ hysteresis measurements were performed mainly on a Vibrational Sample Magnetometer (VSM) operating at 40 Hz with 3 mm oscillation amplitude in a Quantum Design Physical Property Measurement System at 300 K. Magnetic measurements of samples, where the total thickness of the magnetic layers was below 5 nm had been probed with the Anomalous Hall Effect (AHE), which is much more sensitive for ultrathin films[23]. The applied current was 10 mA. From the measured hysteresis curves, the uniaxial magnetic anisotropy constants $K_U$ were determined as described in one of our previous studies[14]. After having subtracted the shape anisotropy from the raw measurements, the easy axis of this anisotropy was perpendicular to the films' surfaces. The in-plane measured hysteresis curves of the single layered films with thicknesses above 50 nm revealed a slight increase of the saturating field instead of a sharp magnetic switching around zero field due to the formation of stripe domains. Their contribution to the magnetic anisotropy and the possible contribution of anisotropy terms of higher order, which may result in a bending of the magnetization curves around the saturating field, are small when compared to the final $K_U$ values and considered in the given error bars. These results will be published elsewhere.

In order to trace the magnetic properties of Fe-Co surfaces, Density Functional Theory (DFT) calculations were performed with the all-electron full potential linearized augmented plane waves method[24] in the generalized gradient approximation[25]. Anisotropy calculations were performed by comparing energies with magnetization along [001] or [100] directions with spin-orbit coupling included in a second-variational approach[26]. The basis set was chosen so that the smallest muffin tin radius ($R$ = 2.31 a.u.) times the largest **k**-vector in the plane waves expansion was set to $RK_{max}$ = 8.5, while a discretization over 140x140x1 = 196000 **k**-vectors was used for numerical integration over the full Brillouin zone. Calculations were performed for a slab geometry with 8 unit cells (17 atomic layers) of bct $Fe_{0.4}Co_{0.6}$ with lattice parameters $a$ = 2.81 Å and $c$ = 2.89 Å, following our previous experimental study[12] and corresponding to $c/a$ = 1.03. Each slab was separated by approximately 13 Å of vacuum. Chemical disorder was treated in the VCA, which has been shown to overestimate the bulk anisotropy in Fe-Co alloys while it correctly describes qualitative trends[5]. The virtual atom closest to the surface shows an enhanced magnetic moment of 2.36 $\mu_B$ while already in the third layer of atoms from the surface, the magnetic moment stays in the range 2.21-2.22 $\mu_B$, which is also the value obtained in a bulk calculation. Hence, the 17 atomic layers are considered to be more than enough to achieve bulk properties in the center. In the VCA, the magnetic moment should be interpreted as the weighted average of Fe and Co moments. Surface anisotropy contributions were finally evaluated by comparing anisotropy energies with a calculation for a bulk bct alloy with same lattice parameters and Brillouin zone integration on a grid of close to 100000 **k**-points (46x46x46).

## Thickness dependence of $K_U$

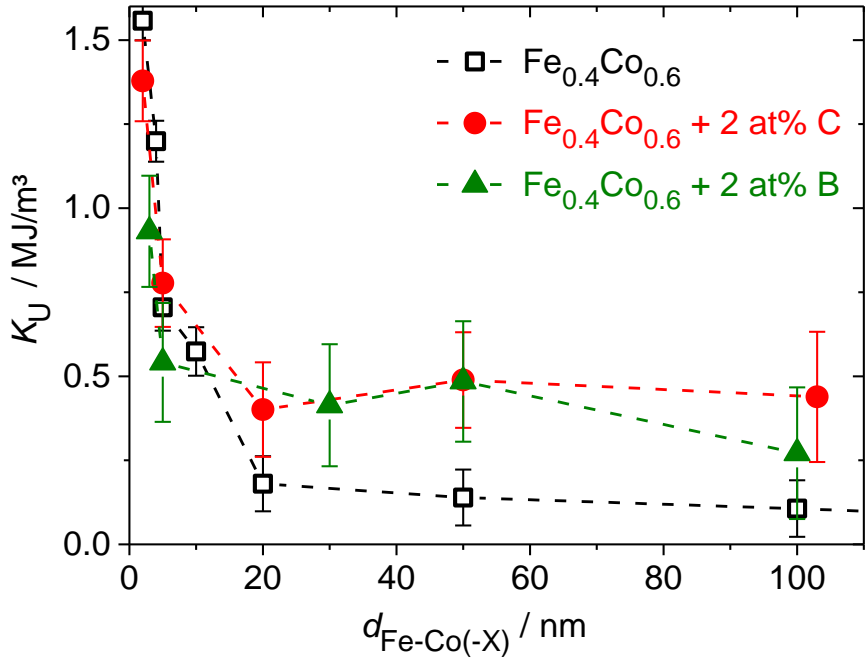

*Fig. 1: Film thickness dependence of the measured perpendicular magnetic anisotropy constants $K_U$ of binary $Fe_{0.4}Co_{0.6}$ films and $(Fe_{0.4}Co_{0.6})_{0.98}X_{0.02}$ films with additions X of B or C. $Au_{0.5}Cu_{0.5}$ buffers were used.*

As shown in previous investigations[14,21], epitaxially grown Fe-Co films with additions of 2 at% B or C exhibit spontaneous strain with $c/a$ of approx. 1.03. Such spontaneous strain is independent from film thickness. As the strained $c$ axis is perpendicular to the films' surface, these films exhibit a uniaxial magnetocrystalline anisotropy. Fig. 1 depicts the measured perpendicular anisotropy constants $K_U$ of Fe-Co(-X) films with different thicknesses. Ultrathin films with thicknesses in the range of few nanometers exhibit the highest magnetic anisotropy approaching the shape anisotropy, which is 1.75 MJ/m³ for the presented films with $\mu_0 M_S$ = 2.1 T. This tendency to huge anisotropies when approaching zero thickness does not depend on whether additions of interstitial elements as B or C have been alloyed to Fe-Co or not. This observation thus indicates a strong composition independent contribution of the buffer interface or the free surface to the magnetic anisotropy. With increasing film thickness, $K_U$ decreases sharply for all Fe-Co(-X) films. In the case of binary $Fe_{0.4}Co_{0.6}$ it almost vanishes (black squares). These films primarily exhibit shape anisotropy, just a very small perpendicular anisotropy remains, which we attribute to the PLD film preparation. The latter is known[27,28] to result in a small strain in film growth direction of 1 %, which is also observed in the binary films presented here. The ternary films containing either 2 at% B or C (red circles and green triangles, respectively), however, exhibit a uniaxial magnetic anisotropy around 0.4 MJ/m³ which remains even in thick films and thus originates from the spontaneous tetragonal strain of their lattices[14,21]. As the whole volume of these films exhibits a strain with $c/a$ between 1.03 and 1.04[14,17,21] and contributions from interfaces or surfaces can be neglected for the higher thicknesses, this value is considered as volume anisotropy $K_{vol}$ of the Fe-Co-X films. Both, the strain and the anisotropy, are independent whether B (green triangles) or C atoms (red circles) were alloyed to Fe-Co, since these atoms both preferentially occupy interstitial sites along the *c* axis[21,22] and thus equivalently contribute to the tetragonal strain. That's why we don't distinguish between Fe-Co-C and Fe-Co-B films with similar stoichiometry (C or B content, respectively) in the present study, using the term Fe-Co-X.

The observed thickness dependence of the magnetic anisotropy agrees well with other experimental studies[7-12], where strong perpendicular anisotropy exclusively was observed in films with thicknesses up to 3 nm. Although these anisotropies were attributed to buffer induced strain, no direct strain measurements were carried out as discussed in the introduction. Before we can assess the strain dependence, we first want to distinguish clearly between the contributions of the volume and (inter- or sur-) faces.

## Separating volume and faces related anisotropy

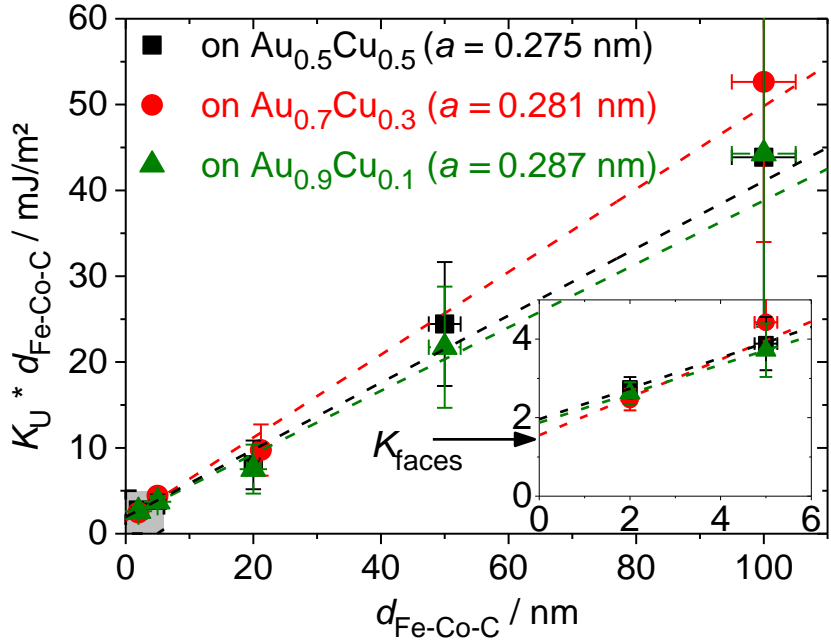

*Fig. 2: Measured perpendicular uniaxial magnetic anisotropy $K_U$ of $(Fe_{0.4}Co_{0.6})_{0.98}C_{0.02}$ films multiplied with their corresponding $(Fe_{0.4}Co_{0.6})_{0.98}C_{0.02}$ film thickness $d_{Fe-Co-C}$ against the thickness. Horizontal error bars represent the accuracy of the film thickness determination. Error bars may be smaller than the data points. The inset depicts the shaded area close to the origin to demonstrate the intersection of the linear fits with the y axis.*

As already indicated in the previous sections, the magnetic properties of thin films are expected to differ from bulk behavior due to their interfaces and surfaces. We will first discuss the sum of the contributions originating from all faces, as this sum is obtained from integral magnetic measurements of single layer films. Since the interface towards the buffer and the free surface may contribute differently, we later will focus on their particular contributions. Following Chappert and Bruno[29], the experimentally accessible total anisotropy $K_U$ is considered as sum of the thickness independent volume anisotropy $K_{vol}$ and the thickness dependent contribution of the faces $K_{faces}$

$K_U = K_{vol} + K_{faces} / d$ (1),

where

$K_{faces} = K_{interface} + K_{surface}$ (2)

with $K_{interface}$ and $K_{surface}$ being the contributions of the buffer interface and the free surface, respectively. Formula (1) may also be written as

$K_U * d = K_{vol} * d + K_{faces}$ (3).

Chappert and Bruno originally suggested this approach to analyze the influence of electronic effects, which occur at interfaces. Clemens et al.[18], however, showed that an identical formula is obtained, when the influence of a continuously relaxing strain is considered. The linear equation (3) allows for a determination of $K_{vol}$ and $K_{faces}$ when the product of the measured $K_U$ and the corresponding film thickness $d$, $K_U * d$, is plotted against $d$. In order to verify the described thickness dependent behavior, such plotting was performed for $(Fe_{0.4}Co_{0.6})_{0.98}C_{0.02}$ single films of different thicknesses, which were deposited on Au-Cu(001) buffer layers with varying composition and thus varying in-plane lattice parameter[15] (Fig. 2). We studied three very different series, where the buffer in-plane lattice parameter $a_{buffer}$ was either smaller (squares), equivalent (circles) or bigger (triangles) when compared to the Fe-Co-X equilibrium lattice parameter[14], which is 0.281 nm. As a consequence, three different strain states should be present at the interface, varying from compressive strain, an unstrained state to tensile strain, respectively. As already discussed, direct reliable strain measurements at the buffer interfaces are very challenging. That's why we focus on the magnetic measurements. The results in Fig. 2 confirm a linear dependence of product $K_U * d$ for the three different Au-Cu buffer series and thus the described model, where the faces related anisotropy depletes with increasing film thickness. From the linear fits, the volume contribution $K_{vol}$ and the interface or surface contribution $K_{faces}$ could be determined for each series: According to eq. (3), the first is the slope of the linear fit, while the latter is its intersection with the y axis, which is enlarged in the inset of Fig. 2.

Since three different buffer lattice parameters with three very different strain states are compared in Fig. 2, a more detailed discussion on the faces related contribution will be done in the following. Therefore, the obtained values, $K_{vol}$ and $K_{faces}$ are plotted against $a_{buffer}$ in Fig. 3. It can be seen that both, the volume and the faces related contribution, do not show a significant dependence on the used buffer layer. As $K_{faces}$ is positive, the interface and the surface contribute to a perpendicular magnetization of films with decreasing thicknesses. We consider the small variations of the slope and the y axis intersection as statistical errors since they do not follow changes of the in-plane lattice parameters of the Au-Cu buffers. The missing dependence of $K_{vol}$ from the applied buffer lattice is in agreement with expectations since the spontaneous strain, which is a volume property, is responsible for this contribution. The observation of buffer independent $K_{faces}$ already indicates that the contribution of a possible induced strain due to the buffer lattice on the magnetic properties is negligible, which confirms our previous results on $(Fe_{0.4}Co_{0.6})_{0.98}C_{0.02}$ films. In Ref.[17], we reported a complete relaxation of the induced strain in films of only 5 nm thickness and no variation of the magnetic anisotropy depending on the used buffer. Again, we argue, that the buffer induced strain relaxes during the film growth and only spontaneous strain remains. TEM observations confirm this statement[14,21], since the c/a ratio measured close to the buffer interface does not vary significantly from the latter. It can be concluded that there is no influence of induced strain on the magnetic properties as no induced strain is left at the time of the magnetic *ex situ* measurements. In contrast to other interface effects, which scale with $1/d$ as described in eq.(1), induced strain should thus not contribute to the magnetic measurements at all. The observation of little magnetic changes in dependence from the buffer lattice here, however, is in contrast to other works, where strong changes of the magnetic anisotropy of Fe-Co films had been observed for films on buffers with different in-plane lattice parameters like Pd[9], Ir[10,11] or Rh[12]. However, one significant aspect is different, when comparing our observations to these experiments from literature: We varied the Au/Cu stoichiometry in $Au_xCu_{1-x}$ to obtain different lattice parameters[15,17]. As both elements, Au and Cu, are within the same column of the periodic table, the number of valence electrons and thus the electrons relevant for the metallic binding, remain constant. This is in contrast to the previous experiments[9-12], where the focus was on the elastic properties instead of the electronic contributions.

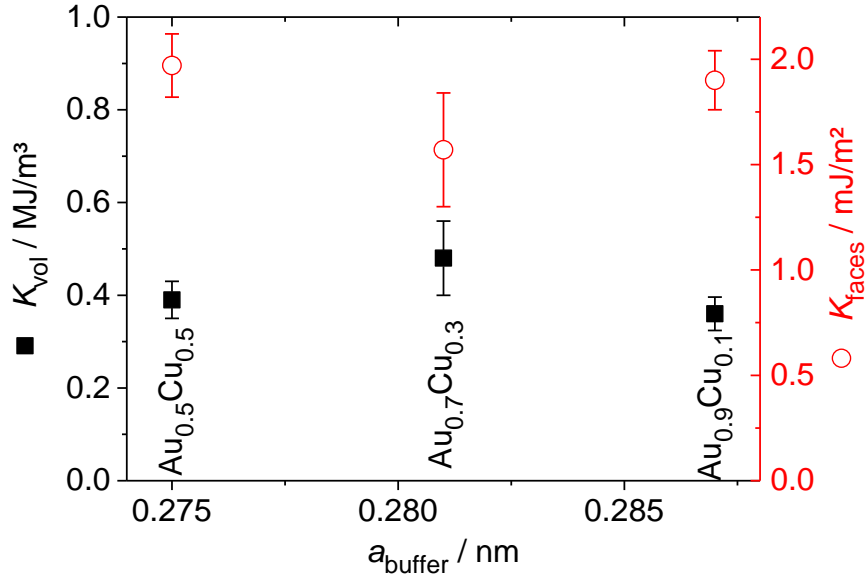

*Fig. 3 Volume (squares, left axis) and faces (circles, right axis) contribution to the measured magnetic anisotropy of $(Fe_{0.4}Co_{0.6})_{0.98}C_{0.02}$ films on different $Au_xCu_{1-x}$ buffers, $K_{vol}$ and $K_{faces}$, respectively.*

As result of this analysis, we obtain an intermediate strain related volume contribution $K_{vol}$ of 0.4 MJ/m³, which can be attributed to the spontaneous strain due to the C interstitial atoms[14,22]. This value agrees with our previous experimental and theoretical study[14]. The sum of the interface and the surface contribution $K_{faces}$ is in the range of 1.8 mJ/m², which is an extraordinary high value. Even when divided by 2, which can be used as a first simple approximation, for an equal contribution of the buffer interface and the free surface, the value of 0.9 mJ/m² outranges the typical interface anisotropies of (001) oriented Fe/Au or Fe/Cu interfaces[30], which are between 0.4 and 0.5 mJ/m². In comparison, the interface anisotropy of Co(001) interfaces with Au or Cu should be even lower, as reviewed by Johnson et al.[31] (and references therein). The only surface anisotropy with a comparable magnitude as observed here, was reported by Urquhart et al.[32] for free Fe(001) surfaces in vacuum with 0.96 mJ/m².

**Separating the contributions of the buffer interface and the free surface**

In single layer films as discussed in the previous section, always both types of interfaces exist: buffer interface and free surface. In experiments, it is impossible to distinguish between the contributions of both faces to the anisotropy. Multilayer films, consisting of alternating layers of the buffer and Fe-Co-X, exclusively exhibit interfaces, but no free surface. In order to determine the contribution of these interfaces to the magnetic anisotropy, multilayers consisting of $(Fe_{0.4}Co_{0.6})_{0.98}B_{0.02}$ layers with thickness $d_{Fe-Co-B}$ on a Au-Cu buffer layers were deposited similarly as the already presented single layered films. Each magnetic layer was covered with a Au-Cu interlayer having the thickness $d_{Au-Cu}$ and the same composition and lattice parameter like the applied buffer. Both layers, $(Fe_{0.4}Co_{0.6})_{0.98}B_{0.02}$ and Au-Cu were repeated *n* times, i.e. that the layer exposed to the surface was again Au-Cu. The epitaxial multilayers thus exclusively have (001) oriented interfaces of type Fe-Co-B/Au-Cu and no free surface. Eq. (1) is thus replaced by

$K_U = K_{vol} + 2\, K_{interface} / d$ (4).

Measurements of the magnetic anisotropy $K_U$ at different thicknesses thus allow for a determination of $K_{interface}$. The factor 2 illustrates the two interfaces of each magnetic layer.

Fig. 4 gives a TEM cross-section of such a sample, having *n* = 10 repetitions, $Au_{0.7}Cu_{0.3}$ buffer and interlayer composition and layer thicknesses $d_{Fe\text{-}Co\text{-}B} = d_{Au\text{-}Cu} = 4$ nm. The TEM analysis revealed a high texture of the complete film architecture and low interface roughnesses. However, XRR measurements and their simulations quantify the latter for the interlayers as between 1 and 2 Å and for the Fe-Co-B layers as between 1 and 8 Å, which may also indicate a certain intermixing of the layers. Similar roughness values were also measured on the other multilayer samples, which will be presented in the following. Based on Bruno's calculations[33,34], Johnson et al.[31] argued that an overestimation of the surface anisotropy by 0.1 mJ/m² can be expected for films with a roughness between 2 and 4 Å, which is low when compared to the estimated 0.9 mJ/m² for each considered face.

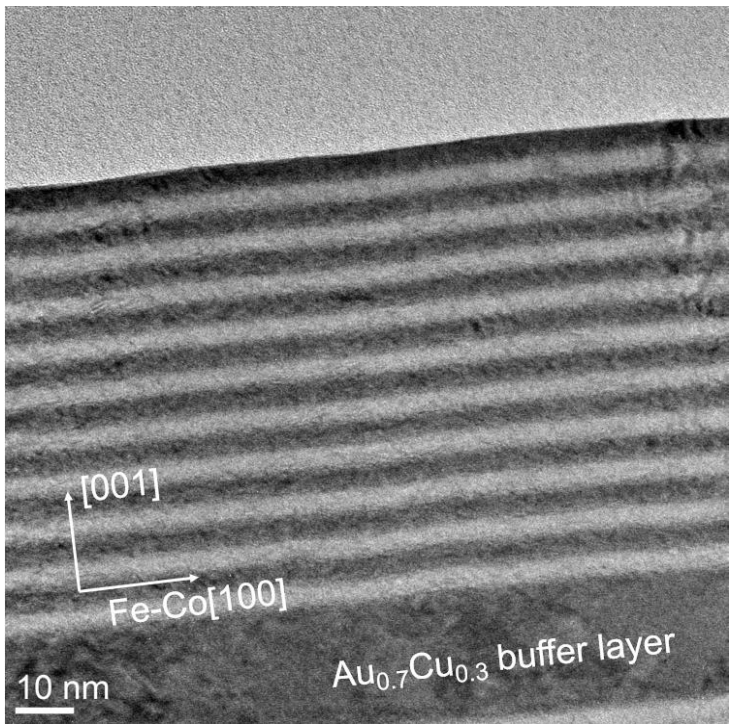

Fig. 4: TEM cross-section of a (4 nm $(Fe_{0.4}Co_{0.6})_{0.98}B_{0.02}$/4 nm $Au_{0.7}Cu_{0.3}$) multilayer with 10 repetitions.

Multilayers similar to the one depicted on Fig. 4 were deposited using different Au-Cu compositions for the buffer and interlayers, respectively. Fig. 5 gives the measured *c/a* ratios (open circles, right axis) of the Fe-Co-B layers and their $K_U$ values (red squares, left axis) versus the lattice parameter of the buffer layer. Within the range of the error bars, both properties remain constant. In particular, the *c/a* ratio within the Fe-Co-B layers does not follow the broken line, which is the expected strain when their lattice would be strained coherently adapting the buffer lattice parameter in the film plane and conserving its volume. Again, we argue that the buffer induced strain is relaxed in the whole films' volumes[17] as only a low strain, which is between 5 and 6 % lattice distortion, remains. Due to the constant strain and film thickness, $K_U$ also remains unaffected by the buffer layers. However, $K_U$ of the multilayers is higher than the previously determined volume anisotropy $K_{vol}$ and thus indicates a strong contribution of interface anisotropy. As this interface anisotropy is independent from the interlayer lattice parameters, we exclude an elastic origin. An electronic origin is very likely and is supported by the observed constant $K_U$ as the density of valence electrons is also considered to be constant in $Au_xCu_{1-x}$ interlayers with different x.

With the intention to increase the influence of the interfaces, epitaxial multilayers with a reduced $(Fe_{0.4}Co_{0.6})_{0.98}B_{0.02}$ layer thickness of 1.5 nm were studied. The thickness of the Au-Cu interlayers was again 4 nm to ensure the (001) texture of the complete multilayers, which consisted of 17 repetitions. RHEED observations, which were performed *in situ* during film growth, revealed a misoriented growth, when the interlayer thickness was reduced.

The perpendicular magnetic anisotropy constants of the multilayers with $d_{Fe-Co-B}$ = 1.5 nm have been added to Fig. 5 (black squares). As observed in the single layered $(Fe_{0.4}Co_{0.6})_{0.98}C_{0.02}$ films, they again do not show any dependence from the Au-Cu lattice parameter. This confirms that the lattice of the buffer or interlayers is not a decisive parameter for $K_U$ of the Fe-Co-B layers, although theoretical works[3-5] predicted a strong dependence of Fe-Co's MCA, when a tetragonal strain would be induced. The maximum anisotropy is expected between *c/a* ratios of 1.2 and 1.25. Due to the low layer thickness and the low film roughness, the present *c/a* ratios could neither be determined *ex situ* with XRD techniques or TEM nor *in situ* with RHEED. However, $K_U$ is strongly increased to values of around 0.95 MJ/m³ in the multilayers with reduced layer thickness, when compared to the multilayers with higher Fe-Co-B layer thickness of averaged 4.2 nm (red squares in Fig. 5). This result confirms the influence of a significant interface anisotropy $K_{interface}$.

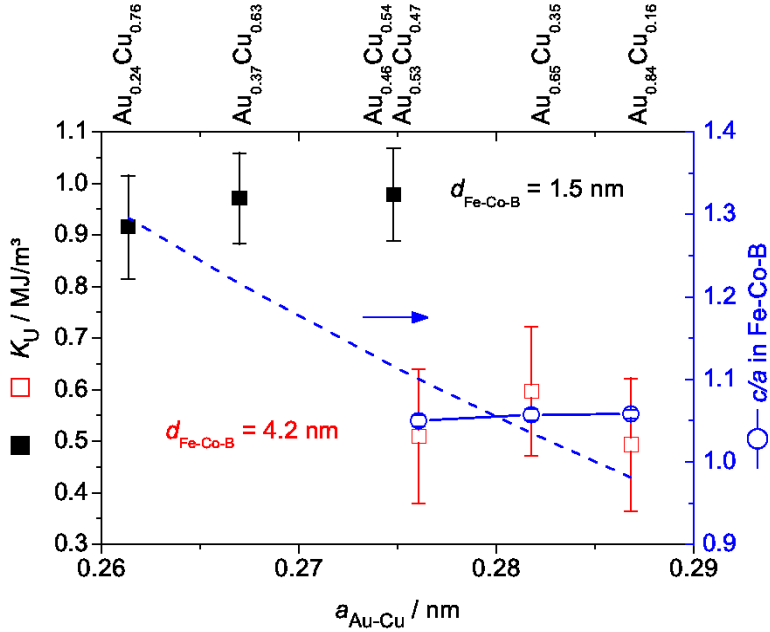

*Fig. 5: Perpendicular uniaxial magnetic anisotropy of multilayer samples with different $(Fe_{0.4}Co_{0.6})_{0.98}B_{0.02}$ layer thicknesses against the lattice parameters of the interlayers (squares). The broken curve gives the expected tetragonal strain c/a according to the epitaxial Bain path[6]. The open circles represent the measured strain in the ML with 4 nm thick $(Fe_{0.4}Co_{0.6})_{0.98}B_{0.02}$ layers.*

Using $K_{vol}$ = 0.4 MJ/m³, which was the result of the studies on single layered films, we calculate $K_{interface}$ as 0.41±0.08 mJ/m² for the multilayers with 1.5 nm Fe-Co-B layer thickness from eq. (4). This value is very close to the 0.38 mJ/m² measured by Warnicke et al.[8] on FeCo/Rh multilayers or estimated 0.34 mJ/m² according to Lao et al.[13] who studied single layered chemically ordered FeCo on Rh. In the latter publication, a value $2K_{interface}$ of 0.68 mJ/m² was calculated, which is the sum of the upper and the lower interface of the regarded films. Compared to the reported values[30,31] for interfaces of Fe or Co to Au or Cu, we observe only a slightly increased anisotropy. We thus argue, that the influence of roughness or interdiffusion in our films is negligible.

The comparison of multilayer and single layer experiments also allows us to determine the anisotropy of the free surface: $K_{surface} = K_{faces} - K_{interface}$ = 1.4 mJ/m². This value by far exceed the anisotropy of the Fe-Co-X/Au-Cu interfaces, which is about 0.4 mJ/m². The very high surface anisotropy, however, is in the same order of magnitude as reported for structurally similar Fe(001), where 0.96 mJ/m² were measured[32].

**Discussing the "free" surface**

The main conclusion of the present experiments is that the substantial contribution to the perpendicular anisotropy of Fe-Co(-X) films originates from the surface and not from the buffer interface. In order to quantify a possible contribution of an ideally flat Fe-Co surface, DFT calculations were performed. A free standing slab with 17 monolayers thickness was modeled therefore. Such slab exhibits a surface anisotropy of only 0.08 mJ/m², when referred to one of the two slab faces. This value is much lower than the one experimentally obtained. It points out, that the present film surface differs strongly from the modeled (ideal) slab geometry. Oxidation, which occurs before all *ex situ* experiments, is very likely to be the origin. Our previous experimental studies of Fe-Co-X films[14,21] indicate that their surface is covered with a thin roughly 2 nm native oxide. Oxygen may have a strong influence on the electron density due to its high electronegativity, which detracts electrons from the Fe-Co-X film. This impact may be stronger, when compared to a metallic interface, which can even add electrons. Compared to the influence of the free surface, the contribution from the Au-Cu interface may be small, but it is still significant (and positive, i.e. benefiting a perpendicular easy axis) as seen in the high $K_U$ values of the multilayer films. An oxidation preventing cap layer like Al may reduce the effect of the free surface, but is certainly again not comparable to an ideal free surface. In our recent study, the volume anisotropy and the spin-orbit coupling in Al covered Fe-Co-B films was studied[35]. There was no striking influence of the faces on the magnetic anisotropy. However, local magnetic measurements on a partially oxidized Fe slab found an increased magnetic orbital to spin ratio on the oxide layer[36] which could lead to increased surface magnetic anisotropy in a uniaxial system like FeCo-X.

Luo et al.[12] tried to separate electronic effects on the magnetic anisotropy from the strain related contribution. The latter was considered as constant within the film's volume, which is a simplified approximation since strain relaxes with increasing film thickness[14,17] and thus contributes similarly to the magnetic anisotropy like an electronic contribution from the interfaces[18]. We speculate that the different anisotropies observed for Fe-Co films on other buffer materials[7-12] like Ir, Pd or Pt may partially have electronic origin and propose to separate the different contributions in future experiments. However, one of our previous studies compared 5 nm thin single layers of $(Fe_{0.4}Co_{0.6})_{0.98}C_{0.02}$ on electronically very different materials, ranging from Ir to MgO, showed only little variation in terms of magnetic anisotropy and indicated that the electronic contribution of the interfaces is low[17]. In accordance with the present study, the electronic contribution from the free oxidized surface may strongly exceed all other interface related anisotropies.

In order to clarify the role of the oxidized surface regarding the magnetic properties, we propose further studies, which go beyond the scope of this report. The influence of the oxide layer could be controlled by a variation of the oxide thickness, crystal structure and orientation. Electrochemical methods allow for a controlled reduction of a present native oxide and reoxidation. *In situ* magnetic measurements[23] should be carried out on ultrathin films. We expect strong variations of the magnetic behavior of such films in dependence of the state of the oxide, similar as has been shown for Fe[37], Co[38], FePt[23,39] and CoPt[40] films. However, the large variety of possible (Fe-Co) surface oxides[41,42] anticipates a very complex behavior.

**Conclusion**

We studied single layered Fe-Co(-X) on Au-Cu buffer layers and multilayers of the type (Fe-Co-X/Au-Cu)$_n$ in order to identify the influence of interface and surface anisotropies. The Fe-Co layers were alloyed with small atoms (X = B or C) on interstitial sites, which resulted in a spontaneous tetragonal strain and a perpendicular magnetocrystalline volume anisotropy around 0.4 MJ/m³ independent from film thickness. This value is in good agreement with previous DFT calculations of strained Fe-Co-X, which considered bulk lattices without any interfaces and hence only computed this volume contribution. Additionally to this contribution, a strong and thickness dependent contribution to the magnetic anisotropy was found and is attributed to electronic effects at both, the free surface and the Au-Cu interfaces. The lattice parameter of the Au-Cu buffer or interlayer has a negligible influence on the magnetic anisotropy of the studied Fe-Co-X films. In other words, no influence of induced strain on the magnetic properties could be proven. As observed in the multilayers, the contribution of the Au-Cu interfaces to the measured perpendicular anisotropy is positive, i.e. favoring perpendicular magnetization, and significant (0.4 mJ/m²). However, the results of single layers without cover layer show that the contribution of the free oxidized surface is much higher (1.4 mJ/m²). The magnetic properties of Fe-Co(-X) films with thicknesses up to around 5 nm are thus strongly determined by the electronic environment at their interfaces. Though a detailed analysis of the presumably oxidized surface is beyond the scope of the present paper, these findings suggest that tailoring the Fe-Co-X/oxide interfaces may open a path towards rare-earth free architectures with high magnetic hardness.

**Acknowledgements**

We acknowledge the EU for funding through FP7-REFREEPERMAG. We thank Tobias Grundmann and Ulrike Besold for experimental support and Karin Leistner for discussion.